\documentclass[prl,aps,superscriptaddress,twocolumn,showpacs]{revtex4-1}

\usepackage{amsfonts}
\usepackage{amsmath}
\usepackage{amssymb}
\usepackage{graphicx}
\usepackage{color}

\begin{document}

\title{Degenerate parametric oscillation in quantum membrane optomechanics}
\author{M\'{o}nica Benito}
\email[]{These authors contributed equally to the project.}
\affiliation{Instituto de Ciencia de Materiales, CSIC, Cantoblanco, 28049 Madrid, Spain}
\affiliation{Max-Planck Institut f\"ur Quantenoptik, Hans-Kopfermann-Str. 1, D-85748 Garching, Germany}
\author{Carlos S\'{a}nchez Mu\~{n}oz}
\email[]{These authors contributed equally to the project.}
\affiliation{F\'{\i}sica Te\'{o}rica de la Materia Condensada, Universidad Aut\'{o}noma de Madrid, 28049 Madrid, Spain}
\affiliation{Max-Planck Institut f\"ur Quantenoptik, Hans-Kopfermann-Str. 1, D-85748 Garching, Germany}
\author{Carlos Navarrete-Benlloch}
\affiliation{Max-Planck Institut f\"ur Quantenoptik, Hans-Kopfermann-Str. 1, D-85748 Garching, Germany}

\begin{abstract}
The promise of innovative applications has triggered the development of many modern technologies capable of exploiting quantum effects. But in addition to future applications, such quantum technologies have already provided us with the possibility of accessing quantum-mechanical scenarios that seemed unreachable just a few decades ago. With this spirit, in this work we show that modern optomechanical setups are mature enough to implement one of the most elusive models in the field of open system dynamics: degenerate parametric oscillation. The possibility of implementing it in nonlinear optical resonators was the main motivation for introducing such model in the eighties, which rapidly became a paradigm for the study of dissipative phase transitions whose corresponding spontaneously broken symmetry is discrete. However, it was found that the intrinsic multimode nature of optical cavities makes it impossible to experimentally study the model all the way through its phase transition. In contrast, here we show that this long-awaited model can be implemented in the motion of a mechanical object dispersively coupled to the light contained in a cavity, when the latter is properly driven with multi-chromatic laser light. We focus on membranes as the mechanical element, showing that the main signatures of the degenerate parametric oscillation model can be studied in state-of-the-art setups, thus opening the possibility of studying spontaneous symmetry breaking and enhanced metrology in one of the cleanest dissipative phase transitions.
\end{abstract}

\maketitle

\textbf{Introduction.} The last decades have seen the birth of a plethora of new technologies working in the quantum regime, starting with the laser \cite{Townes,Basov,Prochorov,Siegman,MilonniEberly,Svelto,NarducciAbraham,WeissVilaseca}, and including nonlinear optics \cite{Bloembergen,Mills,Boyd,BlueBook,DrummondHillery}, trapped ions \cite{Dehmelt,Paul,Wineland,LeibfriedRev,PorrasRev} and atoms \cite{Chu,Cohen-Tannoudji,Phillips,Ketterle,CornellWieman,MetcalfStraten,Ashkin,JakschZollerRev,BlochDalibardZwergerRev}, cavity quantum electrodynamics \cite{Haroche,RaimondRev,KimbleRev,WaltherRev}, or, more recently, superconducting circuits \cite{DevoretRev12,GirvinRev08,ClarkeRev08,NoriRev05,DevoretRev04} and optomechanical resonators \cite{AspelmeyerRev14,MarquardtRev09,KippenbergRev07}. Apart from their potential for quantum computation \cite{NielsenChuang,Bacon10,Smith12} and simulation \cite{Feynman60,Feynman82,Lloyd96,CiracZoller,BlochQsimRev,BlattQsimRev,AspuruGuzikQsimRev,KochQsimRev}, quantum metrology \cite{Lloyd04,Lloyd11}, and quantum communication \cite{Gisin02,Gisin07,CerfBook}, all these technologies have allowed us to reach physical scenarios that were nothing but a dream (or a `gedanken' experiment) for the founding fathers of quantum mechanics.

\begin{figure*}[t]
\begin{center}
\includegraphics[width=\textwidth]{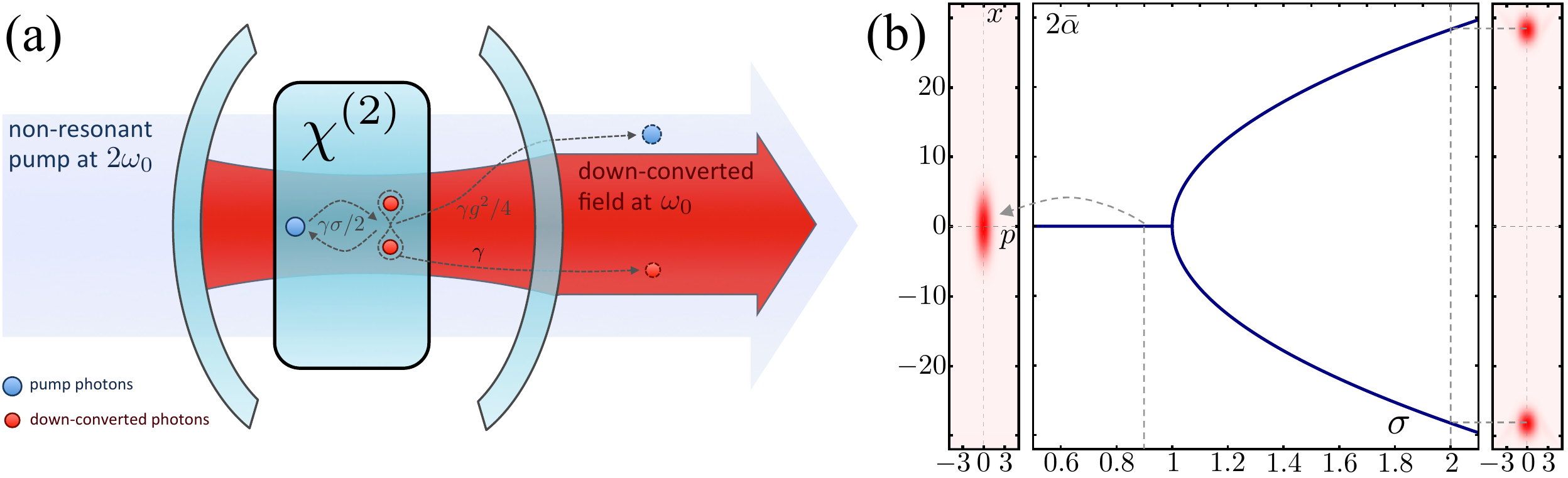}
\end{center}
\caption{(a) Sketch of the optical implementation of the DPO model. (b) Phase transition of the DPO model. In the central panel we plot the steady-state field amplitude as a function of $\sigma$, as predicted by the classical theory. The phase transition is revealed as non-analytic behaviour of the amplitude at threshold, and the presence of two stable solutions above threshold connected by the symmetry transformation $\hat{a}\rightarrow -\hat{a}$. The left and right panels show the steady-state Wigner function for $\sigma=0.9$ and $2$, respectively, which for our purposes here can be interpreted as the joint probability density function $W(x,p)$ providing the statistics of `position' ($\hat{x}=\hat{a}^\dagger+\hat{a}$) and `momentum' ($\hat{p}=\mathrm{i}\hat{a}^\dagger-\mathrm{i}\hat{a}$) measurements. It can be appreciated that even though the quantum steady state is unique and preserves the sign symmetry for any $\sigma$, it does so in two very distinct ways above or below threshold: with a single state centered in phase space in the first case, and with a mixture of two symmetry-breaking states above threshold, which deep into the phase-transition tend to the coherent states predicted by the classical limit.}
\label{fig:1}
\end{figure*}

In this work we keep deepening into the possibility of using new technologies to access phenomena that, even though predicted and theoretically analyzed since decades ago, have eluded observation so far, or only until very recently \cite{Leghtas15}. In particular, we show how modern optomechanical setups based on oscillating membranes \cite{Thompson08,Jayich08,Zwickl08,Wilson09,Sankey10,Vitali12,Purdy12,Schnabel12,Vitali13a,Vitali13b,Harris15} allow for the implementation of \textit{degenerate parametric oscillation} (DPO), a fundamental model in the field of dissipative phase transitions \cite{HakenBook,NicolisBook,LugiatoBook,Drummond80c,Drummond80q}. Together with the laser, DPO has possibly the best-studied quantum-optical dissipative model, since it holds the paradigm of a phase transition whose associated spontaneously broken symmetry is discrete \cite{Drummond80c,Drummond80q} (in contrast to that of the laser, which is continuous \cite{Haken70,Scully70}). Even though the main motivation for studying such a model came from nonlinear optics during the eighties \cite{BlueBook}, in particular from the possibility of implementing it in optical parametric oscillators, see Fig. \ref{fig:1}, the intrinsic multi-mode nature of optical cavities prevents its implementation above the phase transition, as we explain in the next section. In other words, despite the great deal of work invested on this model and its optical implementation, degenerate \textit{optical} parametric oscillators (DOPOs) do not exist in reality.

The situation is rather different in the microwave realm of electronic circuits, where one can build single-mode cavities in the form of simple LC circuits. Indeed, it is in this context where DPO has been traditionally studied in more detail all the way through its phase transition \cite{Kawakubo73,Kawakubo78,Kawakubo79}. However, an electronic circuit at room temperature has a very strong thermal microwave background which, together with other sources of technical noise, completely masks quantum noise and hence any possibility of analyzing quantum mechanical effects. This scenario was radically changed with the advent of superconducting circuits \cite{DevoretRev12,GirvinRev08,ClarkeRev08,NoriRev05,DevoretRev04}, which are cooled down to mK temperatures, effectively removing the thermal background and making it possible to access the quantum regime. It is in this scenario where, just a few months ago, quantum mechanical effects appearing as one crosses the phase transition of the DPO model have been finally observed \cite{Leghtas15}.

Apart from being a clean system where studying fundamental questions related to spontaneous symmetry breaking and ergodicity of open quantum systems \cite{Graham,Cresser,Molmer97,Molmer97b}, DPO might serve as a perfect test bed for enhanced metrology via dissipative phase transitions \cite{Kasia14,Guta14,Molmer14,Guta11}. Motivated then by the interest that this model generates on different communities ranging from the purely theoretical to the most applied ones, in this work we show that DPO can be implemented in the motion of a membrane dispersively coupled to the field of an optical cavity \cite{Thompson08,Jayich08,Zwickl08,Wilson09,Sankey10,Vitali12,Purdy12,Schnabel12,Vitali13a,Vitali13b,Harris15}, when a multi-chromatic laser properly drives the latter. Starting from a first principles model, analytical and numerical methods allow us to identify the regimes where the desired model appears, as well as proving the feasibility of the scheme for the parameters under which current experiments take place.

\textbf{Degenerate parametric oscillation and its optical implementation.}
Let us start by introducing the DPO model, taking its optical implementation as the guiding context (Fig. \ref{fig:1}), and discussing some of the physics derived from it. In its minimal formulation a DOPO consists of an optical cavity containing a crystal with second order nonlinearity, and pumped by a laser at frequency $2\omega_0$ for which the cavity is transparent (non-resonant pump configuration). The parametric down-conversion process occurring inside the $\chi^{(2)}$ crystal is able to generate photons at the subharmonic frequency $\omega_0$, assumed resonant. The model can then be formulated as the following master equation for the state $\hat{\rho}$ of the intracavity field \cite{Kinsler91,CarmichaelBook2}:
\begin{equation}
\frac{d\hat{\rho}}{dt}=-\mathrm{i}\left[\hat{H}_\mathrm{DPO},\hat{\rho}\right]+\frac{\gamma g^2}{4}\mathcal{D}_{a^2}[\hat{\rho}]+\gamma\mathcal{D}_a[\hat{\rho}],
\label{MasterDPO}
\end{equation}
with
\begin{equation}
\hat{H}_\mathrm{DPO}=\omega_0\hat{a}^\dagger\hat{a}+\mathrm{i}\gamma\sigma(e^{-2\mathrm{i}\omega_0 t}\hat{a}^{\dagger2}-e^{2\mathrm{i}\omega_0 t}\hat{a}^2)/2,
\label{HamDPO}
\end{equation}
where $\hat{a}$ is the annihilation operator of cavity photons, and we use the notation $\mathcal{D}_J[\hat{\rho}]=2\hat{J}\hat{\rho}\hat{J}^\dagger-\hat{J}^\dagger\hat{J}\hat{\rho}-\hat{\rho}\hat{J}^\dagger\hat{J}$. The last term describes the loss of cavity photons through the partially transmitting mirror (with corresponding damping rate $\gamma$, proportional to the mirror transmittance). The second term describes the loss of photon pairs which, after being down-converted into a pump photon, leave the cavity to never come back (at rate $\gamma g^2/4$, where $g$ is proportional to the crystal's nonlinear susceptibility). Finally, the Hamiltonian term describes the coherent exchange of photon pairs with the pumping field via down-conversion (at rate $\gamma\sigma/2$, where $\sigma$ is proportional to the amplitude of the laser), as well as the free evolution of the cavity mode.

The first thing to note is that this master equation is invariant under the parity transformation $\hat{U}=(-1)^{\hat{a}^\dagger\hat{a}}$, which performs the operation $\hat{U}^\dagger\hat{a}\hat{U}=-\hat{a}$. On the other hand, defining $\bar{\alpha}=\lim_{t\rightarrow\infty}e^{\mathrm{i}\omega_0 t}\langle\hat{a}(t)\rangle$, it is well known \cite{BlueBook,NavarretePhD,SupMat} that the classical limit of this equation predicts an \textit{off} (or \textit{below-threshold}) stationary state $\bar{\alpha}=0$ for $\sigma\leq 1$, and an \textit{on} (or \textit{above-threshold}) phase-bistable state $\bar{\alpha}=\pm\sqrt{2(\sigma-1)}/g$ for $\sigma>1$ \cite{SupMat}. Hence, at $\sigma=1$ (\textit{threshold}) the classical theory predicts a phase transition, accompanied by spontaneous symmetry breaking of the discrete phase above threshold, since the system has to choose between two possible steady states which individually do not preserve the symmetry.

\begin{figure*}[t]
\begin{center}
\includegraphics[width=\textwidth]{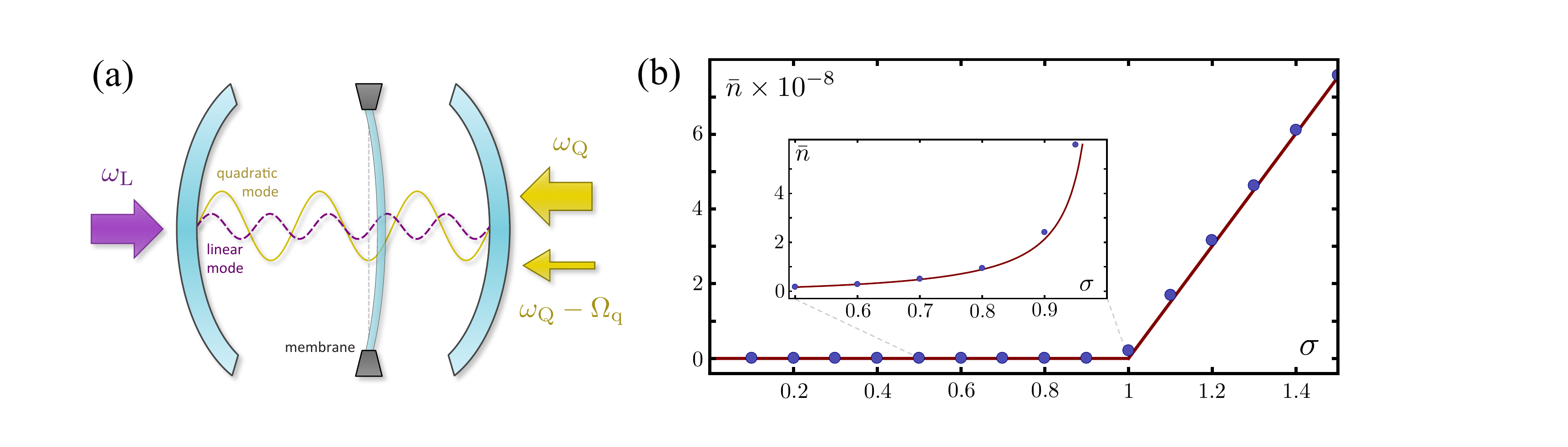}
\end{center}
\caption{(a) Sketch of the optomechanical setup with which we propose to implement the DPO model. (b) Asymptotic phonon number as a function of the pump parameter, for a set of parameters in correspondence with current experiments (see the text for details). The main plot corresponds to the predictions in classical limit, and the inset to the predictions obtained from the asymptotic state generated by the master equation. The blue circles correspond to the optomechanical model, while the red curves are evaluated from its effective DPO model. Note the good agreement between both models.}
\label{fig:2}
\end{figure*}

In contrast to the classical state, the quantum steady-state solution $\bar{\rho}=\lim_{t\rightarrow\infty}e^{\mathrm{i}\omega_0 t\hat{a}^\dagger\hat{a}}\hat{\rho}(t)e^{-\mathrm{i}\omega_0 t\hat{a}^\dagger\hat{a}}$ of Eq. (\ref{MasterDPO}) is unique for any $\sigma$ \cite{Drummond80q,Carmichael88,CarmichaelBook2}. The symmetry of the master equation, together with the uniqueness of the steady state, forces the latter to be invariant under the transformation as well, $\hat{U}\bar{\rho}\hat{U}^\dagger=\bar{\rho}$, what in turn implies that $\bar{\alpha}=\mathrm{tr}\{\bar{\rho}\hat{a}\}=0\;\;\forall\sigma$. This could lead to the conclusion that quantum theory completely spoils the phase transition and its associated spontaneous symmetry breaking. However, the situation is a bit more subtle: as shown in Fig. \ref{fig:1}b through the Wigner function \cite{SupMat,NavarreteNumericsNotes,Weedbrook12,NavarreteQInotes,Navarrete14PRL,Cahill69,Brune92,Garraway92}, below threshold the quantum state is a squeezed state centered at the origin of phase space, while above threshold it develops two lobes centered (approximately) around the classical bi-stable solutions. This shows that in the quantum domain the phase transition predicted at the classical level has the significance of a crossover between phases which preserve the symmetry in two physically distinct ways.

Unfortunately, in real experiments degenerate down-conversion has to compete with non-degenerate channels in which the photon pairs are generated in two different modes with frequencies $\omega_1$ and $\omega_2$ such that $\omega_1+\omega_2=2\omega_0$ (\textit{energy conservation}), and it is possible to show that phase-matching in the nonlinear crystal (\textit{momentum conservation}) always gives preference to one of such processes above threshold \cite{Harris69,Eckardt91,Fabre97}. Therefore, these devices cannot be used to study experimentally the DPO model all the way through its phase transition.

\textbf{Optomechanical implementation of degenerate parametric oscillation.} In contrast to the optical case, we show in this work that optomechanical resonators in which a mechanical degree of freedom is dispersively coupled to the cavity field allow for the implementation of DPO all the way through its phase transition. Our proposal follows closely current experimental setups based on dielectric membranes embedded in optical cavities \cite{Thompson08,Jayich08,Zwickl08,Wilson09,Sankey10,Vitali12,Purdy12,Schnabel12,Vitali13a,Vitali13b,Harris15}. In such setups, the type of coupling arising between the membrane's motion and a given driven cavity mode depends on the position of the former with respect to the standing wave defined in the cavity by the latter \cite{Thompson08,Jayich08,Sankey10,Vitali13a}. In particular, denoting by $\hat{x}$ the displacement of the membrane with respect to its equilibrium position (normalized to its zero-point fluctuations \cite{AspelmeyerRev14}), the frequency shift felt by the optical mode is proportional to $\hat{x}^2$ when the membrane is located in a node or an antinode of the mode's standing wave, while it is proportional to $\hat{x}$ when it is half-way between them. In most optomechanical systems the linear coupling dominates, and many exciting phenomena have been already proven by exploiting it, including mechanical cooling \cite{Gigan06,Arcizet06,Schliesser06,Corbitt07,Thompson08,Wilson09,Teufel11,Chan11,Vitali12,WilsonRae07,Marquardt07,Genes08,Harris15}, optical squeezing \cite{Fabre94,Mancini94,Brooks12,Safavi13},  and induced transparency \cite{Weis10,Safavi11omit,Teufel11omit,Massel12,Vitali13b}. On the other hand, the quadratic coupling has already promised very interesting applications such as quantum nondemolition measurements of the phonon number \cite{Thompson08,Jayich08}, and in our case it will provide the leading mechanism to achieve degenerate parametric oscillation.

Our proposal is sketched in Fig. \ref{fig:2}a. We consider two optical modes with frequencies $\omega_\mathrm{l}$ and $\omega_\mathrm{q}$, linearly and quadratically coupled to fundamental mechanical mode of the membrane (with frequency $\Omega$), respectively; consequently, we will refer to them as the \emph{linear} and \emph{quadratic} modes. This can be achieved with a single optical cavity by selecting two different resonances with the appropriate relative separation between nodes and antinodes. In the absence of optomechanical coupling, the membrane is at some equilibrium temperature $T$ with its thermal environment, which drives it at some rate $\gamma_\mathrm{m}$ to a thermal state with mean phonon number $\bar{n}_\mathrm{th}=k_B T/\hbar\Omega$. However, we assume the linear mode to be driven by a monochromatic laser at frequency $\omega_\mathrm{L}$ tuned to cool down the membrane to an effective phonon number $\bar{n}_\mathrm{eff}\approx\bar{n}_\mathrm{th}/C_\mathrm{l}$ at rate $\gamma_\mathrm{eff}=C_\mathrm{l}\gamma_\mathrm{m}$, where $C_\mathrm{l}$ is the cooperativity of the linear optomechanical coupling \cite{WilsonRae07,Marquardt07}; current experiments reach cooperativities on the order of 10000, which together with cryogenic temperatures allow to cool down the mechanical motion close to its ground state \cite{Harris15} $\bar{n}_\mathrm{eff}=0$, what we assume in the following. On the other hand, the quadratic mode is driven with a laser containing a tone at frequency $\omega_{\mathrm{Q}}$, plus a sideband at frequency $\omega_{\mathrm{Q}}-\Omega_\mathrm{q}$. We will use the first tone to create the two-phonon losses needed in the DPO model (\ref{MasterDPO}), while the combined action of the two tones will provide the coherent exchange of phonon pairs.

We model the system by a master equation governing the evolution of its state $\hat{\rho}$, which in a frame rotating at the laser frequency $\omega_\mathrm{Q}$ takes the form
\begin{equation}
\frac{d \hat \rho}{dt} = -\mathrm{i}[\hat{H}_\mathrm{m}+\hat{H}_\mathrm{q}(t)+\hat{H}_\mathrm{qm},\hat{\rho}]+\gamma_\mathrm{q}\mathcal{D}_{a_\mathrm{q}}[\hat{\rho}]+\gamma_\mathrm{eff}\mathcal{D}_b[\hat{\rho}],
\label{OMmasterEq}
\end{equation}
with Hamiltonian terms
\begin{subequations}
\begin{eqnarray}
\hat{H}_\mathrm{m}&=&\Omega\hat{b}^\dagger\hat{b},
\\
\hat{H}_\mathrm{q}&=&\Delta_\mathrm{q}\hat{a}_\mathrm{q}^\dagger\hat{a}_\mathrm{q}+\mathrm{i}[\mathcal{E}_\mathrm{q}(t)\hat{a}_\mathrm{q}^\dagger-\mathcal{E}_\mathrm{q}^*(t)\hat{a}_\mathrm{q}],
\\
\hat{H}_\mathrm{qm}&=&-g_\mathrm{q}\hat{x}^2\hat{a}_\mathrm{q}^\dagger\hat{a}_\mathrm{q},
\end{eqnarray}
\end{subequations}
where the bichromatic driving amplitude of the quadratic mode can be written as $\mathcal{E}_\mathrm{q}(t)=\mathcal{E}_0+\mathcal{E}_1 e^{-\mathrm{i}\Omega_\mathrm{q}t+\mathrm{i}\phi}$, with $\phi$ some relative phase between the two tones. $\Delta_\mathrm{q}=\omega_\mathrm{Q}-\omega_\mathrm{q}$ is the laser detuning. $\hat{b}$ is the mechanical annihilation operator, from which the mechanical displacement is written as $\hat{x}=\hat{b}+\hat{b}^\dagger$. $\hat{a}_\mathrm{q}$ is the quadratic mode's annihilation operator, with corresponding cavity damping rate $\gamma_\mathrm{q}$ (proportional to the mirror transmissivity). The driving amplitudes $\mathcal{E}_{0,1}$ can be written in terms of the power $P_{0,1}$ of the laser at the corresponding frequency as $\mathcal{E}_{0,1}\approx\sqrt{2\gamma_\mathrm{q}P_{0,1}/\hbar\omega_\mathrm{q}}$ \cite{NavarretePhD}. 

We can understand the conditions under which this bichromatically-driven optomechanical model is mapped to the DPO model by adiabatically eliminating the quadratic mode. We provide the details of the derivation in \cite{SupMat}, and here we just want to point out some physically relevant steps. We follow the usual projector-superoperator technique \cite{ZollerBook,BreuerPetruccione,CarmichaelBook1,Degenfeld15} in which the quadratic mode is assumed to be in some reference state and follow some reference dynamics. For our current purposes, it is enough to assume that it does not feel any mechanical backaction, so that its dynamics is described by the master equation above with $g_\mathrm{q}=0$. This means that: (i) we can take a coherent state with amplitude
\begin{equation}
\alpha_\mathrm{q}(t)=e^{\mathrm{i}\hspace{0.3mm}\mathrm{arctan}(\Delta_\mathrm{q}/\gamma_\mathrm{q})}[\sqrt{\bar{n}_0}-\mathrm{i}\sqrt{\bar{n}_1}e^{-\mathrm{i}\Omega_\mathrm{Q}t}]
\end{equation}
as its reference state, where we have made a concrete choice of the second tone's phase $\phi$ that simplifies the expression \cite{SupMat}, and defined $\bar{n}_j=\mathcal{E}_j^2/[\gamma_\mathrm{q}^2+(\Delta_\mathrm{q}+j\Omega_\mathrm{q})^2]$, which are interpreted as the number of photons introduced by the corresponding laser in the cavity; and (ii) all the correlation functions of its quantum fluctuations $\delta\hat{a}_\mathrm{q}=\hat{a}_\mathrm{q}-\alpha_\mathrm{q}$ will decay in time at rate $\gamma_\mathrm{q}$.

Keeping in mind the traditional picture of sideband cooling \cite{WilsonRae07,Marquardt07,Genes08}, it is intuitive to understand how the DPO model arises from $\hat{H}_\mathrm{qm}$ upon adiabatic elimination of the optical mode. First, it will turn out to be convenient to work in the weak sideband regime $\bar{n}_1\ll\bar{n}_0$ \cite{SupMat}. The coherent part of the optical field generates then an effective mechanical Hamiltonian
\begin{eqnarray}
\hat{H}_\mathrm{eff}&=&\Omega\hat{b}^\dagger\hat{b}-g_\mathrm{q}|\alpha_\mathrm{q}(t)|^2\hat{x}^2
\\
&\approx&\Omega_\mathrm{eff}\hat{b}^\dagger\hat{b}+\mathrm{i}g_\mathrm{q}\sqrt{\bar{n}_0\bar{n}_1}(e^{\mathrm{i}\Omega_\mathrm{q}t}\hat{b}^2-\mathrm{H.c.}), \nonumber
\end{eqnarray}
where $\Omega_\mathrm{eff}=\Omega-2g_\mathrm{q}\bar{n}_0$, and any other term can be neglected as long as we work within the rotating-wave approximation $\Omega_\mathrm{eff}\gg g_\mathrm{q}\bar{n}_0$ \cite{SupMat}. Provided that the second tone is chosen as $\Omega_\mathrm{q}=2\Omega_\mathrm{eff}$, this effective Hamiltonian provides precisely $\hat{H}_\mathrm{DPO}$ as introduced in (\ref{HamDPO}). On the other hand, the elimination of the optical fluctuations generates both  dissipative and Hamiltonian mechanical terms. Within the weak sideband and rotating-wave approximations, all the Hamiltonian terms can be neglected, while only the dissipators $\mathcal{D}_{b^2}$, $\mathcal{D}_{b^{\dagger 2}}$, and $\mathcal{D}_{b^\dagger b}$, survive, corresponding, respectively, to two-phonon cooling, heating, and dephasing. In the weak sideband regime, the rate of each process is static and solely controlled by the fundamental tone. Hence, setting its detuning to the red two-phonon resonance, $\Delta_\mathrm{q}=-2\Omega_\mathrm{eff}$, while working in the resolved sideband regime, $4\Omega_\mathrm{eff}^2\gg\gamma_\mathrm{q}^2$, the heating and dephasing terms are highly suppressed \cite{SupMat} just as in standard cooling \cite{WilsonRae07,Marquardt07,Genes08}, leaving us with a two-photon cooling dissipator at rate $\gamma_\mathrm{m}C_\mathrm{q}$, where we have introduced the cooperativity $C_\mathrm{q}=g_\mathrm{q}^2\bar{n}_0/\gamma_\mathrm{q}\gamma_\mathrm{m}$. It is important to note that the Markov approximation (central to this method), requires a decay of the optical correlators much faster than the effective mechanical dynamics induced by the optical fields \cite{SupMat}, that is, $\gamma_\mathrm{q}\gg\mathrm{max}\{\gamma_\mathrm{eff},\gamma_\mathrm{m}C_\mathrm{q},g_\mathrm{q}\sqrt{\bar{n}_0\bar{n}_1}\}$.

This shows that the dynamics of the mechanical mode should follow an effective DPO master equation of the form (\ref{MasterDPO}), with $\omega_0=\Omega_\mathrm{eff}$, $\gamma=\gamma_\mathrm{eff}$, $g^{2}=4C_\mathrm{q}/C_\mathrm{l}$, and $\sigma=g\sqrt{\bar{n}_1\gamma_\mathrm{q}/\gamma_\mathrm{eff}}$. Remarkably, we obtain a set of parameters that can be optically tuned by means of the two laser powers $P_0$ and $P_1$ and that, as we will show below, allow to explore physical regimes not available in all-optical implementations. In the following we discuss whether the parameters which can be reached in current experimental setups, together with the bounds that the above-mentioned conditions impose on these, are compatible with having a reasonable range for these effective DPO parameters.

\textbf{Implementability in current setups.} We take the experiments of \cite{Harris15} as a reference, in which $\Omega=4.4\mathrm{MHz}$, $\gamma_\mathrm{m}=0.8\mathrm{Hz}$, $\omega_\mathrm{q}=1770\mathrm{THz}$, $\gamma_\mathrm{q}=1.3\mathrm{MHz}$, $C_\mathrm{l}=10000$, and $g_\mathrm{q}=10^{-5}\mathrm{Hz}$ (this last parameter taken from \cite{Vitali13a}). In order to stay safely within the rotating-wave approximation we must impose a bound on the intracavity photon number given by $\bar{n}_0\ll\Omega/2g_\mathrm{q}\approx 10^{11}$, what also certifies that we are within the resolved sideband regime, $\Omega_\mathrm{eff}/\gamma_\mathrm{q}\approx 3.5$. Taking $\bar{n}_0=3.3\times10^9$ (corresponding to a power of $P_{0}\approx20\mathrm{mW}$, fairly reasonable in such setups), we then obtain a quadratic cooperativity $C_\mathrm{q}\approx 3.3\times 10^{-7}$, leading to an effective two-phonon loss $g\approx 10^{-5}$, on the order of the one obtained in optical implementations \cite{NavarretePhD}. On the other hand, let's take $\bar{n}_1=3\times 10^8\ll\bar{n}_0$ (corresponding to $P_1\approx 35\mathrm{\mu W}$) as an upper bound for the sideband photon number; then by varying the sideband power from zero to this value, the effective $\sigma$ parameter can be varied all the way through the phase transition and up to $\sigma=2.5$, showing how current optomechanical setups should be able to reach regions of the DPO model beyond what's possible in optical implementations. Note finally that the Markov approximation is very well satisfied since $\gamma_\mathrm{q}/\gamma_\mathrm{eff}\approx 150$.

\textbf{Numerical simulations.} In order to certify the predictions offered above with the effective mechanical model under optical adiabatic elimination, we have performed numerical simulations of the full optomechanical problem for the realistic parameters of the previous section. We proceed in two ways. First, by projecting the master equation (\ref{OMmasterEq}) in a truncated Fock basis for the optical and mechanical modes, allowing us to directly simulate the evolution starting from any initial state \cite{SupMat,NavarreteNumericsNotes}. Since the master equation is manifestly time-dependent, so will be the asymptotic state; in particular, we are interested in the asymptotic phonon number $\lim_{t\rightarrow\infty}\langle\hat{b}^\dagger\hat{b}(t)\rangle$, which oscillates at frequency $\Omega_\mathrm{q}$ and can be approximately written as $\bar{n}+\delta n \sin(\Omega_\mathrm{q} t)$, where typically $\bar{n}\gg\delta n$. In the inset of Fig. \ref{fig:2}b we show the static phonon background $\bar{n}$ as a function of the effective $\sigma$ (which can be tuned through the sideband power $P_1$ keeping the rest of parameters fixed), together with the phonon number predicted from the effective DPO master equation (\ref{MasterDPO}). For large phonon numbers this type of ``brute force'' simulation becomes unfeasible, and does not allow us to get above $\sigma\approx 0.95$. Fortunately, in order to prove that the DPO phase transition is present in the full model, it is enough to consider the classical limit of the model, which is the second type of simulation that we have performed. In this limit the optical and mechanical modes are described by complex amplitudes $\alpha$ and $\beta$, respectively, which, defining the mechanical position $x=2\mathrm{Re}\{\beta\}$ and momentum $p=2\mathrm{Im}\{\beta\}$, evolve according to \cite{SupMat} 
\begin{subequations}
\begin{align}
\dot{x}& =\Omega p,
\\
\dot{p}& =-2\gamma_\mathrm{eff} p-(\Omega-4g_\mathrm{q}|\alpha|^2)x,
\\
\dot{\alpha}& =\mathcal{E}_\mathrm{q}(t)-(\gamma_\mathrm{q}-\mathrm{i}\Delta_\mathrm{q}-\mathrm{i}g_\mathrm{q}x^2)\alpha.
\end{align}
\end{subequations}
These are a set of coupled nonlinear equations which can be efficiently simulated in virtually all parameter space. They predict an asymptotic phonon number given by $\lim_{t\rightarrow\infty}[x^2(t)+p^2(t)]/4$, the static part of which we plot as a function of $\sigma$ in Fig. \ref{fig:2}b. We also show in the figure the predictions of the DPO model in the classical limit, which are analytical and given by $\bar{n}=2(\sigma-1)/g^2$. We see that both the quantum and classical simulations find very good agreement with the DPO model, in particular the classical limit, which shows the phase transition exactly as expected.  

\textbf{Conclusions.} In summary, we have shown that the elusive degenerate parametric oscillator model can be realistically implemented in current optomechanical setups. Apart from providing the possibility of studying experimentally many interesting theoretical predictions put forward during the last three decades, the implementation of this simple (but paradigmatic) dissipative model in modern quantum technologies opens the way to analyzing open questions related to ergodicity and spontaneous symmetry breaking, as well as enhanced metrology with dissipative phase transitions.

\begin{acknowledgements}
\textbf{Acknowledgements}. Our work has benefited from discussions with many colleagues, including Chiara Molinelli, Tao Shi, Yue Chang, Alejandro Gonz\'alez-Tudela, Germ\'an J. de Valc\'arcel, and J. Ignacio Cirac. M.B. and C.S.M. thank the theory division of the Max-Planck Institute of Quantum Optics for their hospitality, as well as G. Platero and C. Tejedor for their respective support. C.N.-B. is most grateful to Prof. Chang-Pu Sun and his group in Beijing, with whom the initial ideas leading to this project were first discussed. M.B. (C.S.M). is supported by the FPI programme of the Spanish MINECO through project MAT2014-58241-P (projects MAT2011-22997 and MAT2014-53119-C2-1-R). C.N.-B. acknowledges funding from the Alexander von Humbolt Foundation through their Fellowship for Postdoctoral Researchers.
\end{acknowledgements}

\newpage
\widetext

\begin{center}
\textbf{Supplemental material}	
\end{center}

This supplemental material is divided in three sections. In the first one, we explain how we obtain the asymptotic state of the degenerate parametric oscillator (DPO) model, both in the quantum and classical regimes. In the second section we make a detailed derivation of the effective mechanical master equation under adiabatic elimination of the optical mode, stressing the conditions and approximations under which it is expected to be equivalent to the DPO model. In the final section we explain how we have found the asymptotic state of the complete optomechanical model, again both in the quantum and classical regimes.

\section{Asymptotic state of the degenerate parametric oscillator}

\textbf{Moving to a simpler picture. }As introduced in the main text, the master equation corresponding to DPO can be formulated for a single bosonic
mode (corresponding to the down-converted intracavity mode in the optical implementation) and it reads 
\begin{equation}\label{MasterDPOsup}
\frac{d\hat{\rho}}{dt}=-\mathrm{i}\left[ \hat{H}_\mathrm{DPO}(t),\hat{\rho}\right] +\frac{\gamma g^2}{4}\mathcal{D}_{a^2}[\hat{\rho}]+\gamma\mathcal{D}_{a}[\hat{\rho}],\end{equation}
with
\begin{equation}\label{HamDPOsup}
\hat{H}_\mathrm{DPO}(t)=\omega_0\hat{a}^\dagger\hat{a}+\mathrm{i}\gamma\sigma(e^{-2\mathrm{i}\omega_0 t}\hat{a}^{\dagger 2}-e^{2\mathrm{i}\omega_0 t}\hat{a}^2)/2,  \end{equation}
where we remind that $\hat{a}$ is the bosonic annihilation operator, and we use the notation $\mathcal{D}_{J}[\hat{\rho}]=2\hat{J}\hat{\rho}\hat{J}^\dagger-\hat{J}^\dagger\hat{J}\hat{\rho}-\hat{\rho}\hat{J}^\dagger\hat{J}$. The parameters are all defined in the main text.

The asymptotic state of the DPO is better analyzed in a picture rotating at frequency $\omega _{0}$, where the state becomes time-independent. In particular, defining the transformation operator $\hat{U}_\mathrm{c}=\exp(\mathrm{i}\omega_0 t\hat{a}^\dagger\hat{a})$, in the new picture the state of the system $\tilde{\rho}=\hat{U}_\mathrm{c}\hat{\rho}\hat{U}_\mathrm{c}^\dagger$ obeys the master equation
\begin{equation}\label{MasterDPOrot}
\frac{d\tilde{\rho}}{dt}=-\mathrm{i}\left[\tilde{H}_\mathrm{DPO},\tilde{\rho}\right] +\frac{\gamma g^2}{4}\mathcal{D}_{a^2}[\tilde{\rho}]+\gamma\mathcal{D}_{a}[\tilde{\rho}],
\end{equation}
where the transformed Hamiltonian reads
\begin{equation}
\tilde{H}_\mathrm{DPO}=\hat{U}_\mathrm{c}\hat{H}_\mathrm{DPO}\hat{U}_\mathrm{c}^\dagger-\omega_0\hat{a}^\dagger\hat{a}=\mathrm{i}\gamma\sigma(\hat{a}^{\dagger 2}-\hat{a}^2)/2.
\end{equation}
Hence, we see how in this picture the master equation becomes time-independent, leading to a stationary asymptotic state of the system.

\textbf{Quantum asymptotic state. }Indeed, the unique steady state of this
master equation is known analytically, in particular in the form of a
positive $P$ distribution \cite{Carmichael88}, from which in principle the elements of the
density operator can be reconstructed in any basis \cite{PositiveP}. However, such a
reconstruction is computationally very demanding, and for our purposes it is
simpler to evaluate the steady state numerically. Concretely, in the main
text we have shown the Wigner function associated to the steady state for $%
g=0.1$ and two different values of the pump parameter, $\sigma =0.9$ and $%
\sigma =2$, above and below the phase transition, respectively. Let us now
spend some time explaining how we have computed this steady states and their
corresponding Wigner functions.

As explained in detail in \cite{NavarreteNumericsNotes}, we perform the numerics by
going to superspace, where the elements of the density matrix in the Fock
basis are gathered in a vector $\vec{\rho}$, and the master equation becomes
then a linear system $d\vec{\rho}/dt=\mathbb{L}_{\mathrm{DPO}}\vec{\rho}$,
where $\mathbb{L}_{\mathrm{DPO}}$ is then a representation of the Liouville
superoperator which induces the DPO dynamics, $\mathcal{L}_{\mathrm{DPO}%
}[\cdot ]=-\mathrm{i}[\tilde{H}_{\mathrm{DPO}},\cdot] +\gamma 
\mathcal{D}_{a}[\cdot ]+\gamma g^{2}\mathcal{D}_{a^{2}}[\cdot ]/4$. Note
that the Fock basis is infinite-dimensional, and hence one has to introduce
a truncation $\{|n\rangle \}_{n=0,1,...,N}$ in order to work in the
computer. In our case, we follow the criterion of truncating to values of $N$
for which the observables we are interested in (e.g., the photon number)
converge up to a three-digit precision or more. The steady state corresponds
then to the eigenvector with zero eigenvalue of $\mathbb{L}_{\mathrm{DPO}}$,
which provides us with the Fock basis components of the steady-state
operator, $\{\bar{\rho}_{mn}\}_{m,n=0,1,...,N}$.

Once we have the density operator in the Fock basis, the Wigner function can
be evaluated as follows. First, given a harmonic oscillator with position
and momentum quadratures, $\hat{x}=\hat{a}^{\dagger }+\hat{a}$ and $\hat{p}=%
\mathrm{i}(\hat{a}^{\dagger }-\hat{a})$, respectively, recall that the
Wigner function $W(x,p)$ can be seen as a joint probability density function
for measurements of such observables, in the sense that the marginal $%
P(x)=\int_{%
%TCIMACRO{\U{211d} }%
%BeginExpansion
\mathbb{R}
%EndExpansion
}dpW(x,p)$ provides the probability density function predicting the
statistics of position measurements (and similarly for the momentum).
Defining the polar coordinates $(r,\varphi )$ in phase space by $%
(x,p)=r(\cos \varphi ,\sin \varphi )$, the Wigner function of the steady
state $\bar{\rho}$ can be found from its components in the Fock basis as 
\cite{Navarrete14PRL,Cahill69,Brune92,Garraway92} 
\begin{equation}
\bar{W}(r,\varphi )=\sum_{mn=0}^{N}\bar{\rho}_{mn}W_{mn}(r,\varphi ),
\label{Wsinglemode}
\end{equation}%
where we have defined the Wigner function of the operator $|m\rangle \langle
n|$, given by 
\begin{equation}
W_{mn}(r,\varphi )=\frac{(-1)^{n}}{\pi }\sqrt{\frac{n!}{m!}}e^{i\varphi
(m-n)}r^{m-n}L_{n}^{m-n}(r^{2})e^{-r^{2}/2},  \label{Wnm}
\end{equation}%
with $L_{n}^{p}(x)$ the modified Laguerre polynomials and where we have
assumed $m\geq n$ (note that $W_{nm}=W_{mn}^{\ast }$).

\textbf{Classical limit and steady state.} Phase transitions in dissipative
systems are usually revealed in the classical limit of the corresponding
quantum models. Let us explain how such limit can be obtained from the
master equation describing the system quantum mechanically. The idea is
indeed quite simple in the case of bosonic systems: the classical limit
consists in assuming that the state of all bosonic modes is coherent, with
an amplitude that will play the role of the classical variable.

In particular, in the case of the single-bosonic mode considered in the DPO,
this means that we assume its state to be a coherent state $|\alpha
(t)\rangle $ at all times, such that the expectation value of any normally
ordered observable factorizes as $\langle \hat{a}^{\dagger k}(t)\hat{a}%
^{l}(t)\rangle =\alpha ^{\ast k}(t)\alpha ^{l}(t)$, where for convenience we
are defining expectation values with respect to the state in the picture
rotating at the laser frequency, that is $\langle \hat{a}^{\dagger k}(t)\hat{%
a}^{l}(t)\rangle =\mathrm{tr}\{\hat{a}^{\dagger k}\hat{a}^{l}\tilde{\rho}%
(t)\}$. This allows us to find an evolution equation for $\alpha (t)$ from
the master equation (\ref{MasterDPOrot}) as follows: given any operator $%
\hat{A}$, the master equation allows us to write the evolution of its
expectation value as 
\begin{equation}
\frac{d\langle \hat{A}\rangle }{dt}=\mathrm{tr}\left\{ \hat{A}\frac{d\tilde{%
\rho}}{dt}\right\} =-\mathrm{i}\langle \lbrack \hat{A},\hat{H}_{\mathrm{DOPO}%
}]\rangle +\frac{\gamma g^{2}}{4}(\langle \lbrack \hat{a}^{\dagger 2},\hat{A}%
]\hat{a}^{2}\rangle +\langle \hat{a}^{\dagger 2}[\hat{A},\hat{a}^{2}]\rangle
)+\gamma (\langle \lbrack \hat{a}^{\dagger },\hat{A}]\hat{a}\rangle +\langle 
\hat{a}^{\dagger }[\hat{A},\hat{a}]\rangle ),
\end{equation}%
which applied to the annihilation operator, and using the coherent-state
ansatz provides us with the classical equation of the DPO%
\begin{equation}
\gamma ^{-1}\dot{\alpha}=\sigma \alpha ^{\ast }-\frac{g^{2}}{2}|\alpha
|^{2}\alpha -\alpha \text{.}  \label{cDPOeq}
\end{equation}%
This is indeed the equation that would have been obtained by using classical
electromagnetic theory on the DOPO, where $\alpha $ would be interpreted as
the normalized amplitude of the optical field. As explained in the text,
this equation has two types of asymptotic, stationary ($\dot{\alpha}=0$)
solutions: the trivial one $\bar{\alpha}=0$, and a nontrivial one $\bar{%
\alpha}=\pm \sqrt{2(\sigma -1)}/g$, which exists only for $\sigma >1$ and
has sign-indeterminacy owed to the symmetry $\alpha \rightarrow -\alpha $ of
Eq. (\ref{cDPOeq}). We use the bar to denote \textquotedblleft stationary
state\textquotedblright\ In order for these solutions to be physical, they
need to be stable against perturbations; their stability can be analyzed by
studying the evolution of small perturbations around them, that is, by
writing $\alpha (t)=\bar{\alpha}+\delta \alpha (t)$, and linearizing Eq. (%
\ref{cDPOeq}) with respect to $\delta \alpha $. Defining the vector $\delta 
\boldsymbol{\alpha }=\mathrm{col}(\delta \alpha ,\delta \alpha ^{\ast })$, one
obtains the linear system $d\delta \boldsymbol{\alpha }/dt=\mathcal{M}\delta 
\boldsymbol{\alpha }$, where the linear stability matrix reads%
\begin{equation}
\mathcal{M}=\left( 
\begin{array}{cc}
-1-2g^{2}|\bar{\alpha}|^{2} & \sigma -g^{2}\bar{\alpha}^{2}/2 \\ 
\sigma -g^{2}\bar{\alpha}^{\ast 2}/2 & -1-2g^{2}|\bar{\alpha}|^{2}%
\end{array}%
\right) .
\end{equation}%
Hence, the stability of a given stationary solution $\bar{\alpha}$ is
determined by the eigenvalues of this matrix: when they all have negative
real part, it will be stable, while if some of them have positive real part,
perturbations will tend to grow, showing that the solution is unstable. In
the case of the trivial solution, the eigenvalues are $\lambda _{\pm }=-1\pm
\sigma $, and hence, it is unstable for $\sigma >1$. On the other hand, the
eigenvalues associated to the nontrivial solution read $\lambda _{\pm
}=-2\sigma +1\pm 1$, which are always negative for $\sigma >1$, and hence
this solution is stable.

Therefore, we see that at the classical level the phase transition is
revealed by a non-analytic change in the stationary solution at threshold $%
\sigma =1$.

\section{Adiabatic elimination of the optical mode}

\textbf{Moving to a simpler picture. }Our starting point is the master
equation of the optomechanical system as we introduced it in the main text: 
\begin{equation}
\frac{d\hat{\rho}}{dt}=-\mathrm{i}[\hat{H}_{\mathrm{m}}+\hat{H}_{\mathrm{q}%
}(t)+\hat{H}_{\mathrm{qm}},\hat{\rho}]+\gamma _{\mathrm{q}}\mathcal{D}_{a_{%
\mathrm{q}}}[\hat{\rho}]+\gamma _{\mathrm{eff}}\mathcal{D}_{b}[\hat{\rho}],
\label{OMmaster}
\end{equation}%
with Hamiltonian terms 
\begin{subequations}
\begin{align}
\hat{H}_{\mathrm{m}}& =\Omega \hat{b}^{\dagger }\hat{b}, \\
\hat{H}_{\mathrm{q}}& =\Delta _{\mathrm{q}}\hat{a}_{\mathrm{q}}^{\dagger }%
\hat{a}_{\mathrm{q}}+\mathrm{i}[\mathcal{E}_{\mathrm{q}}(t)\hat{a}_{\mathrm{q%
}}^{\dagger }-\mathcal{E}_{\mathrm{q}}^{\ast }(t)\hat{a}_{\mathrm{q}}], \\
\hat{H}_{\mathrm{qm}}& =-g_{\mathrm{q}}\hat{a}_{\mathrm{q}%
}^{\dagger }\hat{a}_{\mathrm{q}}\hat{x}^{2},
\end{align}%
and where the bichromatic driving amplitude of the quadratic mode can be
written as $\mathcal{E}_{\mathrm{q}}(t)=\mathcal{E}_{0}+\mathcal{E}_{1}e^{-%
\mathrm{i}\Omega _{\mathrm{q}}t+\mathrm{i}\phi }$, with $\phi $ some
relative phase between the two tones that will be chosen shortly. All the
symbols have the meaning introduced in the main text.

In order to perform the adiabatic elimination of the optical mode, it is
convenient to move to a picture where the large coherent background that the
driving fields create in the optical mode is already taken into account.
This is accomplished by using a displacement $\hat{D}[\alpha _{\mathrm{q}%
}(t)]=\exp [\alpha _{\mathrm{q}}(t)\hat{a}_{\mathrm{q}}^{\dag }-\alpha _{%
\mathrm{q}}^{\ast }(t)\hat{a}_{\mathrm{q}}]$ as the transformation operator,
where the amplitude $\alpha _{\mathrm{q}}(t)$ is chosen to obey the
evolution equation 
\end{subequations}
\begin{equation}
\dot{\alpha}_{\mathrm{q}}=\mathcal{E}_{\mathrm{q}}(t)-(\gamma _{\mathrm{q}}-%
\mathrm{i}\Delta _{\mathrm{q}})\alpha _{\mathrm{q}}\text{,}
\end{equation}%
with solution%
\begin{equation}
\alpha _{\mathrm{q}}(t)=\alpha _{\mathrm{q}}(0)e^{-(\gamma _{\mathrm{q}}-%
\mathrm{i}\Delta _{\mathrm{q}})t}+\frac{\mathcal{E}_{0}}{\gamma _{\mathrm{q}%
}-\mathrm{i}\Delta _{\mathrm{q}}}\left[ 1-e^{-(\gamma _{\mathrm{q}}-\mathrm{i%
}\Delta _{\mathrm{q}})t}\right] +\frac{\mathcal{E}_{1}e^{\mathrm{i}\phi }}{%
\gamma _{\mathrm{q}}-\mathrm{i}(\Delta _{\mathrm{q}}+\Omega _{\mathrm{q}})}%
\left[ e^{-\mathrm{i}\Omega _{\mathrm{q}}t}-e^{-(\gamma _{\mathrm{q}}-%
\mathrm{i}\Delta _{\mathrm{q}})t}\right] .
\end{equation}%
Note that choosing $\phi =-\pi /2+\arctan (\Delta _{\mathrm{q}}/\gamma _{%
\mathrm{q}})-\arctan [(\Delta _{\mathrm{q}}+\Omega _{\mathrm{q}})/\gamma _{%
\mathrm{q}}]$, we obtain the asymptotic displacement%
\begin{equation}
\lim_{t\gg \gamma _{\mathrm{q}}^{-1}}\alpha _{\mathrm{q}}(t)=e^{\mathrm{i}%
\hspace{0.3mm}\mathrm{arctan}(\Delta _{\mathrm{q}}/\gamma _{\mathrm{q}})}[%
\sqrt{\bar{n}_{0}}-\mathrm{i}\sqrt{\bar{n}_{1}}e^{-\mathrm{i}\Omega _{%
\mathrm{q}}t}],
\end{equation}%
with $\bar{n}_{0}=\mathcal{E}_{0}^{2}/[\gamma _{\mathrm{q}}^{2}+\Delta _{%
\mathrm{q}}^{2}]$ and $\bar{n}_{1}=\mathcal{E}_{1}^{2}/[\gamma _{\mathrm{q}%
}^{2}+(\Delta _{\mathrm{q}}+\Omega _{\mathrm{q}})^{2}]$, which is the
amplitude that we introduced in the main text. In the following we assume to
be working in this asymptotic regime $t\gg \gamma _{\mathrm{q}}^{-1}$, even
if we don't write the limit explicitly to shorten the expressions. In this
new picture, the transformed state $\tilde{\rho}=\hat{D}^{\dag }[\alpha _{%
\mathrm{q}}(t)]\hat{\rho}\hat{D}[\alpha _{\mathrm{q}}(t)]$ evolves then
according to%
\begin{equation}
\frac{d\hat{\rho}}{dt}=-\mathrm{i}[\tilde{H}(t),\hat{\rho}]+\gamma _{\mathrm{%
q}}\mathcal{D}_{a_{\mathrm{q}}}[\hat{\rho}]+\gamma _{\mathrm{eff}}\mathcal{D}%
_{b}[\hat{\rho}]  \label{OMmasterDisplaced}
\end{equation}%
where the transformed Hamiltonian%
\begin{equation}
\tilde{H}(t)=\hat{D}^{\dag }[\alpha _{\mathrm{q}}(t)][\hat{H}_{\mathrm{m}}+%
\hat{H}_{\mathrm{q}}(t)+\hat{H}_{\mathrm{qm}}]\hat{D}[\alpha _{\mathrm{q}%
}(t)]+\mathrm{i}\left( \dot{\alpha}_{\mathrm{q}}^{\ast }\hat{a}_{\mathrm{q}}-%
\dot{\alpha}_{\mathrm{q}}\hat{a}_{\mathrm{q}}^{\dag }\right) ,
\end{equation}%
can be written as the sum of three terms, $\tilde{H}=\tilde{H}_{\mathrm{q}}+%
\tilde{H}_{\mathrm{m}}(t)+\tilde{H}_{\mathrm{qm}}(t)$, with 
\begin{subequations}
\begin{align}
\tilde{H}_{\mathrm{q}}& =-\Delta _{\mathrm{q}}\hat{a}_{\mathrm{q}}^{\dagger }%
\hat{a}_{\mathrm{q}}, \\
\tilde{H}_{\mathrm{m}}(t)& =\Omega \hat{b}^{\dag }\hat{b}-g_{\mathrm{q}%
}|\alpha _{\mathrm{q}}(t)|^{2}\hat{x}^{2}, \\
\tilde{H}_{\mathrm{qm}}(t)& =-g_{\mathrm{q}}[\alpha _{\mathrm{q}}^{\ast }(t)%
\hat{a}_{\mathrm{q}}+\alpha _{\mathrm{q}}(t)\hat{a}_{\mathrm{q}}^{\dag }]%
\hat{x}^{2}-g_{\mathrm{q}}\hat{a}_{\mathrm{q}}^{\dagger }\hat{a}_{%
\mathrm{q}}\hat{x}^{2}.
\end{align}

The interest of moving to this picture is that now the driving has been
moved to the coupling and the mechanical Hamiltonians, what will allow us to
easily understand the physics behind the system. Moreover, in the absence of
optomechanical coupling the dynamics of the optical mode is generated by the
Liouvillian $\mathcal{L}_{\mathrm{q}}[\cdot ]=\mathrm{i}[\Delta _{\mathrm{q}}%
\hat{a}_{\mathrm{q}}^{\dagger }\hat{a}_{\mathrm{q}},\cdot ]+\gamma _{\mathrm{%
q}}\mathcal{D}_{a_{\mathrm{q}}}[\cdot ]$, including only detuning and
dissipation, which drives it to a vacuum state at rate $\gamma _{\mathrm{q}}$%
---which in the original picture corresponds to a coherent state with
amplitude $\alpha _{\mathrm{q}}(t)$---.

\textbf{Derivation of the effective mechanical master equation. }Let us
rewrite the master equation (\ref{OMmasterDisplaced}) as 
\end{subequations}
\begin{equation}
\frac{d\hat{\rho}}{dt}=\mathcal{L}_{\mathrm{m}}^{(t)}[\hat{\rho}]+\mathcal{L}%
_{\mathrm{q}}[\hat{\rho}]+\mathcal{L}_{\mathrm{qm}}^{(t)}[\hat{\rho}],
\end{equation}%
where $\mathcal{L}_{\mathrm{q}}$ is defined above, while $\mathcal{L}_{%
\mathrm{m}}^{(t)}[\cdot ]=-\mathrm{i}[\tilde{H}_{\mathrm{m}}(t),\cdot
]+\gamma _{\mathrm{eff}}\mathcal{D}_{b}[\cdot ]$ and $\mathcal{L}_{\mathrm{qm%
}}^{(t)}[\cdot ]=-\mathrm{i}[\tilde{H}_{\mathrm{qm}}(t),\cdot ]$. Adiabatic
elimination proceeds by choosing some reference state and dynamics for the
optical mode, which is assumed to remain unperturbed by the mechanical mode,
and hence the accuracy of the elimination depends crucially on the choice of
a proper reference. For our purposes, it is enough to take the dynamics
generated by $\mathcal{L}_{\mathrm{q}}$ as the optical reference, and hence
its steady state $\bar{\rho}_{\mathrm{q}}=|0\rangle _{\mathrm{q}}\langle 0|$
(vacuum) as the reference state.\ Let us then define the projector
superoperator
\begin{equation}
\mathcal{P}[\cdot]= \bar{\rho}_{\mathrm{q}}\otimes\mathrm{tr}_\mathrm{q}\{\cdot\},
\end{equation}
and its complement $\mathcal{Q}=1-\mathcal{P}$. These superoperators satisfy
the useful relations $\mathcal{PL}_{\mathrm{m}}^{(t)}[\cdot ]=\mathcal{L}_{%
\mathrm{m}}^{(t)}\mathcal{P}[\cdot ]$ (obvious since $\mathcal{L}_{\mathrm{m}%
}^{(t)}$ acts on the mechanics only) and $\mathcal{PL}_{\mathrm{q}}[\cdot
]=0=\mathcal{L}_{\mathrm{q}}\mathcal{P}[\cdot ]$ (where the second equality
is again obvious, while the first one comes from $\mathcal{L}_{\mathrm{q}}$
being traceless by conservation of probability).

The next step consists on projecting the master equation onto the
corresponding subspaces defined by these superoperators. Defining the
projected components of the density operator $\hat{u}(t)=\mathcal{P}[\hat{%
\rho}(t)]$ and $\hat{w}(t)=\mathcal{%
Q[}\hat{\rho}(t)]$, and using the properties of the projectors, it is
straightforward to get the coupled linear system 
\begin{subequations}
\begin{align}
\frac{d\hat{u}}{dt}& =\left( \mathcal{L}_{\mathrm{m}}^{(t)}+\mathcal{PL}_{%
\mathrm{qm}}^{(t)}\right) [\hat{u}]+\mathcal{PL}_{\mathrm{qm}}^{(t)}[\hat{w}%
], \\
\frac{d\hat{w}}{dt}& =\left( \mathcal{L}_{\mathrm{m}}^{(t)}+\mathcal{L}_{%
\mathrm{q}}+\mathcal{PL}_{\mathrm{qm}}^{(t)}\right) [\hat{w}]+\mathcal{QL}_{%
\mathrm{qm}}^{(t)}[\hat{u}].
\end{align}%
The second equation can be formally integrated, leading to 
\end{subequations}
\begin{equation}
\hat{w}(t)=\int_{0}^{t}dt^{\prime }\mathcal{T}\left\{ e^{\int_{t^{\prime
}}^{t}dt^{\prime \prime }\left( \mathcal{L}_{\mathrm{m}}^{(t^{\prime \prime
})}+\mathcal{L}_{\mathrm{q}}+\mathcal{PL}_{\mathrm{qm}}^{(t^{\prime \prime
})}\right) }\right\} \mathcal{QL}_{\mathrm{qm}}^{(t^{\prime })}[\hat{u}%
(t^{\prime })],
\end{equation}%
where $\mathcal{T}$ is the time-ordering superoperator, and we have not
written the term which depends on the initial value $\hat{w}(0)$ since in
this concrete dissipative scenario any information related to it has to be
completely washed out asymptotically. Next, we introduce this formal
solution in the first equation, and perform a Born approximation in which we
neglect terms beyond quadratic order in the interaction $\mathcal{L}_{\mathrm{qm}}$%
, what allows us to write%
\begin{equation}
\frac{d\hat{u}}{dt}=\mathcal{L}_{\mathrm{m}}^{(t)}[\hat{u}]+\mathcal{PL}_{%
\mathrm{qm}}^{(t)}[\hat{u}]+\int_{0}^{t}d\tau \mathcal{PL}_{\mathrm{qm}%
}^{(t)}\mathcal{U}_{\mathrm{m}}^{(t-\tau ,t)}e^{\mathcal{L}_{\mathrm{q}}\tau
}\mathcal{QL}_{\mathrm{qm}}^{(t-\tau )}[\hat{u}(t-\tau )],  \label{Ueq}
\end{equation}%
where in addition we have made the variable change $t^{\prime }=t-\tau $ in
the integral, used the fact that $\mathcal{L}_{\mathrm{m}}$ and $\mathcal{L}%
_{\mathrm{q}}$ commute, and defined the mechanical time evolution
superoperator%
\begin{equation}
\mathcal{U}_{\mathrm{m}}^{(t-\tau ,t)}=\mathcal{T}\left\{ e^{\int_{t-\tau
}^{t}dt^{\prime}\mathcal{L}_{\mathrm{m}}^{(t^{\prime
})}}\right\} .
\end{equation}

After performing the partial trace over the optical mode, this equation
provides an effective master equation for the reduced mechanical state $\hat{%
\rho}_{\mathrm{m}}=\mathrm{tr}_{\mathrm{q}}\{\hat{\rho}\}$. In order to
simplify further such equation, we need to write the explicit form of the
interaction $\mathcal{L}_{\mathrm{qm}}$, what we do as%
\begin{equation}
\mathcal{L}_{\mathrm{qm}}^{(t)}[\cdot ]=\mathrm{i}\sum_{j=1}^{3}G_{j}(t)[\hat{B}_j\otimes\hat{x}^2,\cdot ],
\end{equation}%
with%
\begin{equation}
\mathbf{\hat{B}}=(\hat{a}_{\mathrm{q}},\hat{a}_{\mathrm{q}}^{\dagger },\hat{a%
}_{\mathrm{q}}^{\dagger }\hat{a}_{\mathrm{q}}),\text{ \ \ and \ \ }\mathbf{G}%
(t)=g_{\mathrm{q}}[\alpha _{\mathrm{q}}^{\ast }(t),\alpha _{\mathrm{q}%
}(t),1].
\end{equation}%
Note the null asymptotic expectation value of the optical operators, $%
\mathrm{tr}_{\mathrm{q}}\{\hat{B}_{j}\bar{\rho}_{\mathrm{q}}\}=0$ $\forall j$%
, meaning that the second term of Eq. (\ref{Ueq}) does not contribute since $%
\mathcal{PL}_{\mathrm{qm}}^{(t)}=0$. Let us then define the optical
correlation functions 
\begin{subequations}
\begin{align}
\mathrm{tr}_{\mathrm{q}}\{\hat{B}_{l}e^{\mathcal{L}_{\mathrm{q}}\tau }[\bar{%
\rho}_{\mathrm{q}}\hat{B}_{j}]\}& =\lim_{t\rightarrow \infty
}\langle \hat{B}_{j}(t)\hat{B}_{l}(t+\tau
)\rangle _{\mathrm{q}}\equiv K_{jl}(\tau ), \\
\mathrm{tr}_{\mathrm{q}}\{\hat{B}_{l}e^{\mathcal{L}_{\mathrm{q}}\tau }[\hat{B%
}_{j}\bar{\rho}_{\mathrm{q}}]\}& =\lim_{t\rightarrow \infty
}\langle \hat{B}_{l}(t+\tau )\hat{B}_{j}(t)\rangle _{\mathrm{q}}\equiv H_{jl}(\tau ),
\end{align}%
where the expectation value refers to the picture rotating at the laser
frequency and we have used the quantum regression theorem \cite{CarmichaelBook1}. Then, the
effective mechanical master equation can be written as 
\end{subequations}
\begin{align}
\frac{d\hat{\rho}_{\mathrm{m}}(t)}{dt}& =\mathcal{L}_{\mathrm{m}}^{(t)}[\hat{%
\rho}_{\mathrm{m}}]+\sum_{jl=1}^{3}G_{j}(t)G_{l}(t)\int_{0}^{t}d\tau
K_{jl}(\tau )\{\hat{x}^{2}\mathcal{U}_{\mathrm{m}}^{(t-\tau ,t)}[\hat{\rho}_{%
\mathrm{m}}(t-\tau )\hat{x}^{2}]-\mathcal{U}_{\mathrm{m}}^{(t-\tau ,t)}[\hat{%
\rho}_{\mathrm{m}}(t-\tau )\hat{x}^{2}]\hat{x}^{2}\}
\label{AsystemNonMarkovianMasterEq} \\
& +\sum_{jl=1}^{N}G_{j}(t)G_{l}(t)\int_{0}^{t}d\tau H_{jl}(\tau )\{\mathcal{U%
}_{\mathrm{m}}^{(t-\tau ,t)}[\hat{x}^{2}\hat{\rho}_{\mathrm{m}}(t-\tau )]%
\hat{x}^{2}-\hat{x}^{2}\mathcal{U}_{\mathrm{m}}^{(t-\tau ,t)}[\hat{x}^{2}%
\hat{\rho}_{\mathrm{m}}(t-\tau )]\}.  \notag
\end{align}

The correlation functions $K_{jl}(\tau )$ and $H_{jl}(\tau )$ can be
evaluated in many ways. Instead of using the dynamics induced by $\mathcal{L}%
_{\mathrm{q}}$ in the Schr\"{o}dinger picture, a particularly simple way of
evaluating them is by using the equivalent quantum Langevin equations of the
optical operators,\ which in the simple case of having dissipation and
detuning only, consist of a single closed equation for the annihilation
operator \cite{ZollerBook}: 
\begin{equation}
\frac{d\hat{a}_{\mathrm{q}}}{dt}=-(\gamma _{\mathrm{q}}-\mathrm{i}\Delta _{%
\mathrm{q}})\hat{a}_{\mathrm{q}}+\sqrt{2\gamma _{\mathrm{q}}}\hat{a}_{%
\mathrm{in}}(t),
\end{equation}%
where the only non-zero input-operator correlators up to fourth order are 
\begin{eqnarray}
&&\langle\hat{a}_{\mathrm{in}}(t_{1})\hat{a}_{\mathrm{in}}^{\dagger
}(t_{2})\rangle =\delta (t_{1}-t_{2}),\text{ \ }\langle \hat{a}_{\mathrm{in}%
}(t_{1})\hat{a}_{\mathrm{in}}^{\dagger }(t_{2})\hat{a}_{\mathrm{in}}(t_{3})%
\hat{a}_{\mathrm{in}}^{\dagger }(t_{4})\rangle =\delta (t_{1}-t_{2})\delta
(t_{3}-t_{4}),\text{\ \ } \\
&&\text{ }\langle \hat{a}_{\mathrm{in}}(t_{1})\hat{a}_{\mathrm{in}}(t_{2})%
\hat{a}_{\mathrm{in}}^{\dagger }(t_{3})\hat{a}_{\mathrm{in}}^{\dagger
}(t_{4})\rangle =\delta (t_{1}-t_{3})\delta (t_{2}-t_{4})+\delta
(t_{1}-t_{4})\delta (t_{2}-t_{3}).  \notag
\end{eqnarray}%
The asymptotic ($t\gg \gamma _{\mathrm{q}}^{-1}$) solution of this equation
reads%
\begin{equation}
\hat{a}_{\mathrm{q}}(t)=\sqrt{2\gamma _{\mathrm{q}}}\int_{0}^{t}dt^{\prime
}e^{-(\gamma _{\mathrm{q}}-\mathrm{i}\Delta _{\mathrm{q}})(t-t^{\prime })}%
\hat{a}_{\mathrm{in}}(t^{\prime }),
\end{equation}%
which together with the correlators of the input operator allows us to write 
\begin{equation}
K_{jl}(\tau )=e^{-(\gamma _{\mathrm{q}}+\mathrm{i}\Delta _{\mathrm{q}})\tau
}\delta _{j1}\delta _{l2},\text{ \ \ and \ \ }H_{jl}(\tau )=e^{-(\gamma _{%
\mathrm{q}}-\mathrm{i}\Delta _{\mathrm{q}})\tau }\delta _{j2}\delta _{l1},
\end{equation}%
which are functions decaying at rate $\gamma _{\mathrm{q}}$.

Hence, of the dynamics induced by the time evolution superoperator $\mathcal{%
U}_{\mathrm{m}}^{(t-\tau ,t)}$, we see that only the processes happening at
a rate faster than or similar to $\gamma _{\mathrm{q}}$ play a role in the
integral terms of the effective mechanical master equation (\ref%
{AsystemNonMarkovianMasterEq}). This brings us to the final major
approximation, known as the Markov approximation: we assume that of all the
processes contributing to the mechanical dynamics, the only term acting
appreciably on the time-scale of the optical decay is the simple oscillation
induced by the term $[\Omega -2g_{\mathrm{q}}(\bar{n}_{0}+\bar{n}_{1})]\hat{b%
}^{\dagger }\hat{b}$ of $\tilde{H}_{\mathrm{m}}$ (we will justify this approximation self-consistently at the end of the derivation). Within this Markov
approximation, we can then approximate 
\begin{equation}
\mathcal{U}_{\mathrm{m}}^{(t-\tau ,t)}[\hat{\rho}_{\mathrm{m}}(t-\tau )\hat{x%
}^{2}]\approx e^{-\mathrm{i}\Omega _{\mathrm{eff}}\tau \hat{b}^{\dagger }%
\hat{b}}[\hat{\rho}_{\mathrm{m}}(t-\tau )\hat{x}^{2}]e^{\mathrm{i}\Omega _{%
\mathrm{eff}}\tau \hat{b}^{\dagger }\hat{b}}\approx \hat{\rho}_{\mathrm{m}%
}(t)\hat{x}(\tau )^{2},
\end{equation}%
where $\Omega _{\mathrm{eff}}=\Omega -2g_{\mathrm{q}}(\bar{n}_{0}+\bar{n}%
_{1})$ and $\hat{x}(\tau )=e^{\mathrm{i}\Omega _{\mathrm{eff}}\tau }\hat{b}%
+e^{-\mathrm{i}\Omega _{\mathrm{eff}}\tau }\hat{b}^{\dagger }$, and similarly%
\begin{equation}
\mathcal{U}_{\mathrm{m}}^{(t-\tau ,t)}[\hat{x}^{2}\hat{\rho}_{\mathrm{m}%
}(t-\tau )]\approx \hat{x}(\tau )^{2}\hat{\rho}_{\mathrm{m}}(t)\text{,}
\end{equation}%
leading to the effective mechanical master equation%
\begin{align}
\frac{d\hat{\rho}_{\mathrm{m}}}{dt}& =\mathcal{L}_{\mathrm{m}}^{(t)}[\hat{%
\rho}_{\mathrm{m}}(t)]+\left\{ [\Gamma (t;\Omega _{\mathrm{eff}})\hat{b}%
^{2}+\Gamma (t;-\Omega _{\mathrm{eff}})\hat{b}^{\dagger 2}+\Gamma (t;0)(2%
\hat{b}^{\dagger }\hat{b}+1)]\hat{\rho}_{\mathrm{m}}\hat{x}^{2}\right. \\
& \left. -\hat{x}^{2}\left[ \Gamma (t;\Omega _{\mathrm{eff}})\hat{b}%
^{2}+\Gamma (t;-\Omega _{\mathrm{eff}})\hat{b}^{\dagger 2}+\Gamma (t;0)(2%
\hat{b}^{\dagger }\hat{b}+1)\right] \hat{\rho}_{\mathrm{m}}+\mathrm{H.c.}%
\right\} ,  \notag
\end{align}%
where, after performing the time integrals (in the asymptotic limit),\ we
have obtained asymptotic time-dependent rates%
\begin{equation}
\Gamma (t;\omega )=\frac{\gamma _{\mathrm{m}}C_{\mathrm{q}}}{1-\mathrm{i}%
(\Delta _{\mathrm{q}}+2\omega )/\gamma _{\mathrm{q}}}\left( 1+\mathrm{i}%
\sqrt{\bar{n}_{1}/\bar{n}_{0}}e^{\mathrm{i}\Omega _{\mathrm{q}}t}\right) +%
\frac{\gamma _{\mathrm{m}}C_{\mathrm{q}}}{1-\mathrm{i}(\Delta _{\mathrm{q}%
}+\Omega _{\mathrm{q}}+2\omega )/\gamma _{\mathrm{q}}}\left( \bar{n}_{1}/%
\bar{n}_{0}-\mathrm{i}\sqrt{\bar{n}_{1}/\bar{n}_{0}}e^{-\mathrm{i}\Omega _{%
\mathrm{q}}t}\right) ,  \label{Rates}
\end{equation}%
with the cooperativity defined as $C_{\mathrm{q}}=g_{\mathrm{q}}^{2}\bar{n}%
_{0}/\gamma _{\mathrm{q}}\gamma _{\mathrm{m}}$.

In order to see that this master equation has all the ingredients that we
need plus many more, so it is just a matter of finding the regime in which
the latter do not contribute, let us rewrite it. By defining the real and
imaginary parts of the rates, $\Gamma _{\mathrm{R}}(t;\omega )=\mathrm{Re}%
\{\Gamma (t;\omega )\}$ and $\Gamma _{\mathrm{I}}(t;\omega )=\mathrm{Im}%
\{\Gamma (t;\omega )\}$, and defining the phonon-number operator $\hat{n}=%
\hat{b}^{\dagger }\hat{b}$, we can write%
\begin{equation}
\frac{d\hat{\rho}_{\mathrm{m}}}{dt}=-\mathrm{i}\left[ \hat{H}_{\mathrm{eff}%
}(t),\hat{\rho}_{\mathrm{m}}\right] +\gamma _{\mathrm{eff}}\mathcal{D}_{b}[%
\hat{\rho}_{\mathrm{m}}]+\Gamma _{\mathrm{R}}(t;\Omega _{\mathrm{eff}})%
\mathcal{D}_{b^{2}}[\hat{\rho}_{\mathrm{m}}]+\Gamma _{\mathrm{R}}(t;-\Omega
_{\mathrm{eff}})\mathcal{D}_{b^{\dagger 2}}[\hat{\rho}_{\mathrm{m}}]+4\Gamma
_{\mathrm{R}}(t;0)\mathcal{D}_{n}[\hat{\rho}_{\mathrm{m}}]+\mathcal{L}_{%
\mathrm{NRW}}^{(t)}[\hat{\rho}_{\mathrm{m}}],
\end{equation}%
where we have defined the effective Hamiltonian $\hat{H}_{\mathrm{eff}}(t)=%
\hat{H}_{\mathrm{DPO}}(t)+\hat{H}_{\mathrm{\perp DPO}}(t)$, containing terms
that we will need for the DPO model%
\begin{equation}
\hat{H}_{\mathrm{DPO}}(t)=\Omega _{\mathrm{eff}}\hat{n}+\mathrm{i}g_{\mathrm{%
q}}\sqrt{\bar{n}_{0}\bar{n}_{1}}(e^{-\mathrm{i}\Omega _{\mathrm{q}}t}\hat{b}%
^{\dagger 2}-e^{\mathrm{i}\Omega _{\mathrm{q}}t}\hat{b}^{2}),
\end{equation}%
plus some that we don't want to contribute%
\begin{align}
\hat{H}_{\mathrm{\perp DPO}}(t)& =[4g_{\mathrm{q}}\sqrt{\bar{n}_{0}\bar{n}%
_{1}}\sin (\Omega _{\mathrm{q}}t)-\Gamma _{\mathrm{I}}(t;\Omega _{\mathrm{eff%
}})+3\Gamma _{\mathrm{I}}(t;-\Omega _{\mathrm{eff}})+4\Gamma _{\mathrm{I}%
}(t;0)]\hat{n} \\
& +[\Gamma _{\mathrm{I}}(t;\Omega _{\mathrm{eff}})+\Gamma _{\mathrm{I}%
}(t;-\Omega _{\mathrm{eff}})+4\Gamma _{\mathrm{I}}(t;0)]\hat{n}^{2}-[g_{%
\mathrm{q}}(\bar{n}_{0}+\bar{n}_{1}+\mathrm{i}\sqrt{\bar{n}_{0}\bar{n}_{1}}%
e^{\mathrm{i}\Omega _{\mathrm{q}}t})\hat{b}^{\dagger 2}+\mathrm{H.c.}],  \notag
\end{align}%
and we have collected into $\mathcal{L}_{\mathrm{NRW}}$ the terms which in
the absence of the sideband are expected not to contribute within the
rotating wave approximation, which read 
\begin{align}
\mathcal{L}_{\mathrm{NRW}}^{(t)}[\hat{\rho}_{\mathrm{m}}]& =[\Gamma
(t;\Omega _{\mathrm{eff}})+\Gamma ^{\ast }(t;-\Omega _{\mathrm{eff}})]\hat{b}%
^{2}\hat{\rho}_{\mathrm{m}}\hat{b}^{2}-\Gamma (t;\Omega _{\mathrm{eff}})\hat{%
b}^{4}\hat{\rho}_{\mathrm{m}}-\Gamma ^{\ast }(t;-\Omega _{\mathrm{eff}})\hat{%
\rho}_{\mathrm{m}}\hat{b}^{4} \\
& +[\Gamma (t;\Omega _{\mathrm{eff}})+\Gamma ^{\ast }(t;0)]\hat{b}^{2}\hat{%
\rho}_{\mathrm{m}}(2\hat{n}+1)-\Gamma (t;\Omega _{\mathrm{eff}})(2\hat{n}+1)%
\hat{b}^{2}\hat{\rho}_{\mathrm{m}}-\Gamma ^{\ast }(t;0)\hat{\rho}_{\mathrm{m}%
}(2\hat{n}+1)\hat{b}^{2}  \notag \\
& +[\Gamma (t;0)+\Gamma ^{\ast }(t;-\Omega _{\mathrm{eff}})](2\hat{n}+1)\hat{%
\rho}_{\mathrm{m}}\hat{b}^{2}-\Gamma (t;0)\hat{b}^{2}(2\hat{n}+1)\hat{\rho}_{%
\mathrm{m}}-\Gamma ^{\ast }(t;-\Omega _{\mathrm{eff}})\hat{\rho}_{\mathrm{m}}%
\hat{b}^{2}(2\hat{n}+1)+\mathrm{H.c..}  \notag
\end{align}

\textbf{Degenerate parametric oscillation regime.} Let's now discuss the
conditions under which the effective mechanical master equation above will
correspond to the master equation of the DPO, Eq. (\ref{MasterDPOsup}).

Looking at the term $\hat{H}_{\mathrm{DPO}}(t)$ of the effective
Hamiltonian, we see that the sideband $\Omega _{\mathrm{q}}$ should be
chosen to match twice the effective mechanical frequency, that is, $2\Omega
_{\mathrm{eff}}$.

On the other hand, we would like the effective two-phonon cooling $\mathcal{D%
}_{b^{2}}$ to dominate over any other irreversible process, in particular
over the two-phonon heating $\mathcal{D}_{b^{\dagger 2}}$ and dephasing $%
\mathcal{D}_{n}$, and to do so with a time-independent rate. Looking at (\ref%
{Rates}), the latter can be naturally accomplished by driving the
fundamental tone much stronger than the sideband, that is, $\bar{n}_{0}\gg 
\bar{n}_{1}$. The static part of the rates (\ref{Rates}) then suggests that
cooling will be enhanced by choosing a detuning of the fundamental driving
tone matching the red two-phonon sideband, $\Delta _{\mathrm{q}}=-2\Omega _{%
\mathrm{eff}}$; indeed, this choice provides the following real, static part
of the rates%
\begin{equation}
\Gamma _{\mathrm{cooling}}=\gamma _{\mathrm{m}}C_{\mathrm{q}},\text{ \ \ }%
\Gamma _{\mathrm{heating}}=\frac{\gamma _{\mathrm{m}}C_{\mathrm{q}}}{%
1+16\Omega _{\mathrm{eff}}^{2}/\gamma _{\mathrm{q}}^{2}},\text{ \ \ and \ \ }%
\Gamma _{\mathrm{dephasing}}=\frac{\gamma _{\mathrm{m}}C_{\mathrm{q}}}{%
1+4\Omega _{\mathrm{eff}}^{2}/\gamma _{\mathrm{q}}^{2}},
\end{equation}%
showing in addition that we need to work in the resolved sideband regime $%
4\Omega_{\mathrm{eff}}^2\gg \gamma _{\mathrm{q}}^2$ in order for heating and
dephasing to be suppressed; in particular, we will define the parameter $%
r=(1+4\Omega _{\mathrm{eff}}^{2}/\gamma _{\mathrm{q}}^{2})^{-1}\ll 1$, which
allows us to approximate the rates by 
\begin{subequations}
\begin{align}
\Gamma (t;\Omega _{\mathrm{eff}})& =\gamma _{\mathrm{m}}C_{\mathrm{q}}\left(
1+\mathrm{i}\sqrt{r}\bar{n}_{1}/\bar{n}_{0}+\mathrm{i}\sqrt{\bar{n}_{1}/\bar{%
n}_{0}}e^{2\mathrm{i}\Omega _{\mathrm{eff}}t}+\sqrt{r(\bar{n}_{1}/\bar{n}%
_{0})}e^{-2\mathrm{i}\Omega _{\mathrm{eff}}t}\right) , \\
\Gamma (t;-\Omega _{\mathrm{eff}})& =-\mathrm{i}\sqrt{r}\gamma _{\mathrm{m}%
}C_{\mathrm{q}}\left( 1+\mathrm{i}\sqrt{\bar{n}_{1}/\bar{n}_{0}}e^{2\mathrm{i%
}\Omega _{\mathrm{eff}}t}-2\mathrm{i}\sqrt{\bar{n}_{1}/\bar{n}_{0}}e^{-2%
\mathrm{i}\Omega _{\mathrm{eff}}t}\right) /2, \\
\Gamma (t;0)& =\gamma _{\mathrm{m}}C_{\mathrm{q}}\left( \bar{n}_{1}/\bar{n}%
_{0}-\mathrm{i}\sqrt{r}+\sqrt{r(\bar{n}_{1}/\bar{n}_{0})}e^{2\mathrm{i}%
\Omega _{\mathrm{eff}}t}-\mathrm{i}\sqrt{\bar{n}_{1}/\bar{n}_{0}}e^{-2%
\mathrm{i}\Omega _{\mathrm{eff}}t}\right) ,
\end{align}%
expressions that together with working in the weak sideband regime, $\bar{n}%
_{1}/\bar{n}_{0}\ll 1$, suggest that the only relevant rate is the real
static part of $\Gamma (t;\Omega _{\mathrm{eff}})$ which provides the
two-phonon cooling rate as desired.

The next constrain on the parameters comes from the fact that there are many
counter-rotating terms that we want not to contribute within the
rotating-wave approximation. Among these, inspection of $\mathcal{L}_{%
\mathrm{NRW}}^{(t)}$ and $\hat{H}_{\mathrm{\perp DPO}}$ shows that the
largest of such rates are $\gamma _{\mathrm{m}}C_{\mathrm{q}}$ and $g_{%
\mathrm{q}}(\bar{n}_{0}+\bar{n}_{1})\approx g_{\mathrm{q}}\bar{n}_{0}$, but
note that $g_{\mathrm{q}}\bar{n}_{0}/\gamma _{\mathrm{m}}C_{\mathrm{q}%
}=\gamma _{\mathrm{q}}/g_{\mathrm{q}}$ which is typically much larger than 1
(optomechanical systems work far from the single-photon strong coupling
regime, specially when referring to the quadratic coupling), and hence $g_{%
\mathrm{q}}\bar{n}_{0}$ is the largest of these two rates. Thus, the
rotating wave-approximation requires $g_{\mathrm{q}}\bar{n}_{0}\ll \Omega _{%
\mathrm{eff}}$. Provided that this approximation holds, we can neglect all
the counter-rotating terms in $\hat{H}_{\mathrm{\perp DPO}}$ and $\mathcal{L}%
_{\mathrm{NRW}}^{(t)}$, approximating them by 
\end{subequations}
\begin{equation}
\hat{H}_{\mathrm{\perp DPO}}\approx -\sqrt{r}\gamma _{\mathrm{m}}C_{\mathrm{q%
}}(11+9\hat{n})\hat{n}/2,
\end{equation}%
and%
\begin{align}
\mathcal{L}_{\mathrm{NRW}}^{(t)}[\hat{\rho}_{\mathrm{m}}]& =\sqrt{\bar{n}%
_{1}/\bar{n}_{0}}\gamma _{\mathrm{m}}C_{\mathrm{q}}e^{2\mathrm{i}\Omega _{%
\mathrm{eff}}t}[\mathrm{i}\hat{b}^{2}\hat{\rho}_{\mathrm{m}}(2\hat{n}+1)-%
\mathrm{i}(2\hat{n}+1)\hat{b}^{2}\hat{\rho}_{\mathrm{m}}+\sqrt{r}\hat{\rho}_{%
\mathrm{m}}(2\hat{n}+1)\hat{b}^{2} \\
& \text{\ \ \ \ \ \ \ \ \ \ \ \ \ \ \ \ \ \ \ \ \ \ \ \ \ \ \ \ \ \ \ \ \ \
\ \ \ }-\sqrt{r}\hat{b}^{2}(2\hat{n}+1)\hat{\rho}_{\mathrm{m}}+\sqrt{r}\hat{%
\rho}_{\mathrm{m}}\hat{b}^{2}(2\hat{n}+1)]+\mathrm{H.c..}  \notag
\end{align}%
This expression clearly shows that $\mathcal{L}^{(t)}_{\mathrm{NRW}}$ is
negligible when compared with the two-phonon cooling term $\gamma _{\mathrm{m%
}}C_{\mathrm{q}}\mathcal{D}_{b^{2}}$. On the other hand, $\hat{H}_{\mathrm{%
\perp DPO}}$ provides a negligible effective mechanical frequency shift, but
also a Kerr term which is expected to be negligible only as long as $\langle 
\hat{n}\rangle \ll \Omega _{\mathrm{eff}}/4.5\sqrt{r}\gamma _{\mathrm{m}}C_{%
\mathrm{q}}$, which puts a bound on the number of phonons. Nevertheless, for
the parameters corresponding to a realistic implementation used in the main
text, we find this bound to be $\sim 10^{13}$, while the classical limit of
the DPO tells us that the phonon number expected at $\sigma =2.5$ is on the
order of $10^{10}$, three orders of magnitude below the limit in which the
Kerr term can start playing a role.

Provided that all these considerations are taken into account, we then
expect the effective mechanical master equation to be very well approximated
by%
\begin{equation}
\frac{d\hat{\rho}_{\mathrm{m}}}{dt}\approx -\mathrm{i}\left[ \hat{H}_{%
\mathrm{eff}}(t),\hat{\rho}_{\mathrm{m}}\right] +\gamma _{\mathrm{eff}}%
\mathcal{D}_{b}[\hat{\rho}_{\mathrm{m}}]+\gamma _{\mathrm{m}}C_{\mathrm{q}}%
\mathcal{D}_{b^{2}}[\hat{\rho}_{\mathrm{m}}],
\end{equation}%
with effective Hamiltonian%
\begin{equation}
\hat{H}_{\mathrm{eff}}(t)\approx \Omega _{\mathrm{eff}}\hat{n}+\mathrm{i}g_{%
\mathrm{q}}\sqrt{\bar{n}_{0}\bar{n}_{1}}(e^{-2\mathrm{i}\Omega _{\mathrm{eff}%
}t}\hat{b}^{\dagger 2}-e^{2\mathrm{i}\Omega _{\mathrm{eff}}t}\hat{b}^{2}),
\end{equation}%
which is exactly the DPO model, Eq. (\ref{MasterDPOsup}).

Note that there is a final constrain on the parameters coming from the
Markov approximation that we performed in the adiabatic elimination: since
we assumed that within the decay of the optical correlators at rate $\gamma
_{\mathrm{q}}$ the only relevant mechanical process is the simple
oscillation at frequency $\Omega _{\mathrm{eff}}$, we need the rates $\gamma
_{\mathrm{eff}}$, $\gamma _{\mathrm{m}}C_{\mathrm{q}}$, and $g_{\mathrm{q}}%
\sqrt{\bar{n}_{0}\bar{n}_{1}}$, to be smaller than $\gamma _{\mathrm{q}}$.
For the parameters considered in the main text, $\gamma _{\mathrm{eff}}$ is
the largest rate of the three (for an $\bar{n}_{1}$ corresponding to $\sigma
=2.5$ in the DPO model or smaller), and it satisfies $\gamma _{\mathrm{eff}%
}/\gamma _{\mathrm{q}}\approx 6\times 10^{-3}$, so we are safely within the
Markov regime.

\section{Asymptotic state of the full model}

Let us in this final section of the supplemental material explain how we
have found the asymptotic state of the optomechanical model numerically. As
with the stationary state of the DPO model, we have performed simulations of
the full master equation (\ref{OMmaster}), as well as simulations in the
classical limit.

In the first case, our starting point has been the optomechanical master
equation in the displaced picture, Eq. (\ref{OMmasterDisplaced}). For
numerical purposes, it is important to work in this displaced picture
because the state of the optical mode should stay close to vacuum, while in
the original picture it is a highly populated coherent state which does not
allow for a reasonable truncation of the optical Fock space. As before, we
set the truncation of the mechanical and optical Fock bases in such a way
that the mean phonon number finds convergence up to the third significant
digit, what typically does not require more than one or two photons in this
displaced picture. The simulation proceeds again as explained in detail in
\cite{NavarreteNumericsNotes}, that is, by moving to superspace where the master
equation (\ref{OMmasterDisplaced}) is turned into a linear system $d\vec{\rho%
}(t)/dt=\mathbb{L}(t)\vec{\rho}(t)$. Note that now the linear problem is
manifestly time-dependent, and therefore, there will be no steady state. In
particular, $\mathbb{L}(t)$ is $2\pi /\Omega _{\mathrm{q}}$-periodic, and
this periodicity is reflected into a time-dependent asymptotic state $\bar{%
\rho}(t)=\lim_{t\rightarrow \infty }\hat{\rho}(t)$, which we find by solving
the linear system numerically starting from different initial conditions $%
\vec{\rho}(0)$, that is, different initial states $\hat{\rho}(0)$. In all
the simulations we have checked that the asymptotic state is independent of
the chosen initial state (e.g., vacuum or the steady state of the DPO for
the mechanics). As explained in the text, the observable we have focused on
is the asymptotic phonon number $\lim_{t\rightarrow \infty }\langle \hat{b}%
^{\dagger }\hat{b}(t)\rangle =\mathrm{tr}\{\hat{b}^{\dagger }\hat{b}\bar{\rho%
}(t)\}$, whose time evolution can be approximated by a function of the type $%
\bar{n}+\delta n\sin (\Omega _{\mathrm{q}}t)$, with $\delta n\ll \bar{n}$.

The superspace simulation of the master equation becomes quite heavy as the
mechanical state gets populated, what has prevented us from performing
simulations for values of the sideband power where the system is expected to
be above the DPO phase transition. Hence, in order to prove that the
optomechanical model leads to the expected phase transition, we have
performed simulations of the optomechanical system in the classical limit.
Similarly to what we did for the DPO model in the first section, this limit
is found by assuming that both the optical and mechanical modes are in a
coherent state at all times. Let us show the procedure explicitly for this
case too. Our starting point is the original optomechanical master equation (%
\ref{OMmaster}), but replacing the mechanical dissipator $\mathcal{D}%
_{b}[\cdot ]$ by $[\hat{x},\{\hat{p},\cdot \}]/2\mathrm{i}$, where $\hat{p}=%
\mathrm{i}(\hat{b}^{\dagger }-\hat{b})$ is the mechanical momentum
quadrature. For high-Q mechanical oscillators which admit a weak-coupling
description of their interaction with the environment, this dissipator leads
to the same physics as the previous one \cite{ZollerBook}, but provides better-looking
classical equations. With this change, the evolution of the expectation
value of any system operator $\hat{A}$ reads 
\begin{equation}
\frac{d\langle \hat{A}\rangle }{dt}=\mathrm{tr}\left\{ \hat{A}\frac{d\hat{%
\rho}}{dt}\right\} =-\mathrm{i}\langle \lbrack \hat{A},\hat{H}_{\mathrm{m}}+%
\hat{H}_{\mathrm{q}}(t)+\hat{H}_{\mathrm{qm}}]\rangle +\gamma _{\mathrm{q}%
}(\langle \lbrack \hat{a}_{\mathrm{q}}^{\dagger },\hat{A}]\hat{a}_{\mathrm{q}%
}\rangle +\langle \hat{a}_{\mathrm{q}}^{\dagger }[\hat{A},\hat{a}_{\mathrm{q}%
}]\rangle )+\frac{\gamma _{\mathrm{eff}}}{2\mathrm{i}}\langle \{[\hat{A},%
\hat{x}],\hat{p}\}\rangle ,
\end{equation}%
where just as with the DPO model, we are defining the expectation value with
respect to the state in the picture rotating at the laser frequency, that
is, $\langle \cdot \rangle =\mathrm{tr}\{\cdot $ $\hat{\rho}\}$, with $\hat{%
\rho}$ the state in the rotating frame. Applied to $\hat{a}_{\mathrm{q}}$, $%
\hat{x}$, and $\hat{p}$, and denoting by $\alpha (t)$ and $\beta (t)$ the
amplitudes of the optical and mechanical coherent states, we find the
classical evolution equations 
\begin{subequations}
\begin{align}
\dot{x}& =\Omega p, \\
\dot{p}& =-2\gamma _{\mathrm{eff}}p-(\Omega -4g_{\mathrm{q}}|\alpha |^{2})x,
\\
\dot{\alpha}& =\mathcal{E}_{\mathrm{q}}(t)-(\gamma _{\mathrm{q}}-\mathrm{i}%
\Delta _{\mathrm{q}}-\mathrm{i}g_{\mathrm{q}}x^{2})\alpha ,
\end{align}
\end{subequations}
with the classical mechanical position and momentum defined as $x=2\mathrm{Re}%
\{\beta \}$ and $p=2\mathrm{Im}\{\beta \}$, respectively. This nonlinear
system can be efficiently simulated numerically for (practically) any
parameter set, and the phonon number that it predicts can be evaluated as $%
\lim_{t\rightarrow \infty }\langle \hat{b}^{\dagger }\hat{b}(t)\rangle
=\lim_{t\rightarrow \infty }[x^{2}(t)+p^{2}(t)]/4$, which again can be
approximated by $\bar{n}+\delta n\sin (\Omega _{\mathrm{q}}t)$, with
(typically) $\delta n\ll \bar{n}$.

\end{document}